\documentstyle{article}
    \LARGE
    \title{Exact Solutions Related to Nonminimal Gravitational Coupling}
    \author{C. M. G. de Sousa \thanks{present address: International Centre of
            Condensed Matter Physics,  Universidade de Bras\'\i lia ,
	    Caixa Postal 04667, \mbox{70919-900, }
            Bras\'\i lia -- DF, Brazil.}
				\, and F. E. Mendon\c ca da Silveira}
    \date{Instituto de F\'\i sica Te\' orica--UNESP, Rua Pamplona 145,
	 01405--900, Bela Vista, \\ S\~ ao Paulo--SP, Brazil}
\begin{document}
\maketitle

\begin{quote}
   {\small {\bf Abstract:} We obtain exact analytic solutions for a typical
	autonomous dynamical system, related to the problem of a vector field
	nonminimally coupled to gravity.}
\end{quote}

$\frac{}{}$

Gravity is presently the only interaction that is not inserted in any of the
unification schemes. Due to this fact, many toy models have appeared
coupling gravity to fields that could have played significant role
in the Early Universe, in such wise as to display desirable
properties. Special interest has been put in gravity coupled
nonminimally to other fields which could generate, among a number of new
effects, a non-singular universe.

In this context, dynamical systems techniques have been applied to solve
the problem of coupling of gravity to a vector field,
whose Lagrangian can be written as $^1$

\begin{equation}
{\cal L}=\sqrt{-g}\left( \frac{R}{k}-\frac{1}{4}F_{\mu \nu}F^{\mu \nu}+
\beta RW_{\mu}W^{\mu}\right) ,
\end{equation}
where $g=\det g_{\mu \nu}$ ($g_{\mu \nu}$ is the metric tensor), $R$ is the
scalar of curvature, $G=k/8\pi $ is the Newtonian gravitational constant
in units that $\hbar =c=1, \mid \beta \mid =1, W_{\mu}$ is an arbitrary
vector field and $F_{\mu \nu} = W_{\left[ \mu ,\nu \right]},$ where the comma
represents ordinary differentiation and the square brackets represent the
skew--symmetric part of $W_{\mu ,\nu}.$

The Lagrangian (1) leads to a set of equations of motion that can be
transformed into an autonomous dynamical system $^1$. By choosing
the Robertson--Walker metric

\[ ds^{2}= dt^{2}-S^{2}(t)\left[d\chi^{2}+\sigma^{2}(\chi )
    (d\theta^{2}+sin^{2}\theta d\phi^{2})\right] , \]
with the ansatz $W^{2}=W^{2}(t)$, the Lagrangian (1) leads to
equations of motion which has a solution given by

\[ W^{2}(t)=\frac{1}{k}\left( 1-\frac{t}{S}\right) , \]
\[ S(t)= (t^{2}+Q^{2})^{1/2} , \]
where $k$ is the Einstein constant and $Q$ is also a contant.
Afterwards one sets $X=3(\dot{S}/S)$ and $Y=(\dot{\Omega}/\Omega ) ,$
where $\Omega =(1/k)+(\beta W^{2}),$ to write the equations of
motion as

\begin{eqnarray}
\dot{X} &=& -\frac{1}{3}X^{2}+XY, \nonumber \\
	& & \\
\dot{Y} &=& -Y^{2}-XY. \nonumber
\end{eqnarray}

Hence, the correct interpretation of functions $X$ and $Y$ is
important to the knowledge of the evolution of the model.

Up to now there was no exact solution for this dynamical system and
the analysis was carried out by using qualitative procedure.
Here we exhibit the explicit solution in a very simple way.

One can rewrite eqs. (2) in polar coordinates,
$\rho =\left( X^{2}+Y^{2}\right)^{1/2}$ and $\theta =\arctan \frac{Y}{X},$
as $^{2,3}$

\begin{eqnarray}
\dot{\rho} &=& -\rho^{2}\left( \frac{1}{3}\cos^{3}\theta +\sin^{3}\theta
+\cos \theta \sin^{2}\theta - \cos^{2}\theta \sin \theta \right) , \nonumber \\
	   & &  \\
\dot{\theta} &=& -2\rho \cos \theta \sin \theta \left( \frac{\cos \theta}{3}
+\sin \theta \right) . \nonumber
\end{eqnarray}

The phase diagram related to this autonomous dynamical system can be obtained
through
\[ \frac{\dot{\rho}}{\dot{\theta}}=\frac{d\rho}{d\theta}, \]
which leads to

\begin{equation}
\rho =C\exp \left( \frac{1}{2}I\right),
\end{equation}
where $C$ is an arbitrary positive constant of integration and

\begin{equation}
I=\int \frac{\frac{1}{3}(\cos^{3}\theta +3\sin^{3}\theta )
+\cos \theta \sin^{2}\theta - \cos^{2}\theta \sin \theta}
{\frac{1}{3}\cos^{2}\theta \sin \theta +\cos \theta \sin^{2}\theta}d\theta .
\end{equation}

After a little algebra, involving only basic trigonometric relations, one can
show that the solution to integral $I$  readily leads to

\begin{equation}
\rho =C \frac {\left| \tan\theta\right| ^{1/2}}
	     {\left| \cos\theta +3\sin\theta \right|}
\end{equation}

We point out that the precise knowledge of the function
$\rho (\theta )$ leads to the sketch of the desired
phase diagram associate with the autonomous dynamical  system
without any qualitative analysis about the dynamical  system
in regions around the origin $^{1,4}.$

On the other hand one can choose to study the dynamical system
qualitatively by means of the projection on the Poincar\' e sphere $^{3,4}$.
This will provide usefull informations concerning the asymptotic
behaviour of the system. As defined in Ref. 3, the Poincar\'e
sphere with unitary radius is placed over the $xy$-plane, this plane
being tangent to the south pole of the sphere. Another frame is in
order and it is placed in the centre of the sphere. Projections onto
the sphere are taken by joining points of the diagram to the centre
of the sphere. This process will gives rise to a drawing on the
sphere which is projected orthogonally on the $xy$-plane. The final
portrait of the phase diagram is, then, in a circle with unitary radius
where the behaviour at infinity is identified with the border of
the circle.

Applying this method $^{3}$ to (2) one can obtain the topologies
around the singular points at infinity, as in Fig.1. The arrows
show the evolution in time. Note that besides the equilibrium points
$\tilde{D}(0,0)$, $\tilde{D}'(0,0)$, $\tilde{C}(0,0)$ and $\tilde{C}'(0,0)$,
there are the points $\tilde{B}(-1/3,0)$ and $\tilde{B}'(1/3,0)$,
the primes indicating antipodes points. The system refuses to give
informations on the topology around the origin by the method of
linearization. Dulac's test shows there are no limiting cycles. Note
that the above mentioned equilibrium points are consistent with eq. (6).

The complete phase diagram which compactifies infinities is shown in Fig.2.
Its shape is very close to that in Ref.4; the difference is due to
the unusual dimension used. The phase diagram in Ref.1 is
obtained in a different way: the projection is taken on the plane
which corresponds to $X=0$ and crosses the poles of the sphere, the
$xy$-plane being tangent to the north pole of the sphere.

Notice that any autonomous planar dynamical system

\begin{eqnarray}
\dot{X} &=& P(X,Y), \nonumber \\
\dot{Y} &=& Q(X,Y). \nonumber
\end{eqnarray}

allows a similar change of variables, and in polar form the phase
diagram arises from

\begin{equation}
  \frac{d\rho}{d\theta}= \rho
  \frac{P(\rho\cos\theta ,\rho\sin\theta )+
           \tan\theta Q(\rho\cos\theta ,\rho\sin\theta )}
       {Q(\rho\cos\theta ,\rho\sin\theta )-
           \tan\theta P(\rho\cos\theta ,\rho\sin\theta )}
\end{equation}
For this reason, exact solutions may be achieved if it is possible
to rewrite (7) as

\[ \frac{d\rho}{d\theta}= R(\rho )\Theta (\theta )  \]

Recently some cosmological models has been
investigated in the so-called multidimensional scenario $^{5}$, and
in a particular case a Maxwell field in higher dimensions is
coupled to Einstein-Hilbert Lagrangian. This also leads to a
qualitative dynamical system analysis, and, as in eq. (7), the
search for exact solutions related to this problem is in progress.

Thanks are due to Professor C. A. P. Galv\~ao  for reading the manuscript
and for helpful suggestions.

\vspace{.3in}

{\Large\bf References}
\begin{description}
   \item[1.] M. Novello and C. Romero, {\it Gen. Rel. Grav.,}
	19 (1987) 1003--1011 (see also references therein)
   \item[2.] W. Hurewicz, {\it Lectures on Ordinary Differential Equations}
	(Dover, New York, 1990).
   \item[3.] A. Andronov {\it et al., Qualitative Theory of Second--Order
	Dynamic Systems} (John Wiley $\&$ Sons, New York, 1973).
   \item[4.] J. Tossa {\it et al., C. R. Acad. Sci. Paris,} 314(1992)339--343.
   \item[5.] J. C. Fabris and J. Tossa, {\it Phys. Lett.\/}B 296
        (1992) 307--310; J. C. Fabris, {\it Gen. Rel. Grav.,} 26 (1994)
135--147.
\end{description}

\vspace{.5in}
{\Large\bf Figures Captions}
\vspace{.5in}

{\bf Figure 1:} Equilibrium points at infinity and the topology imposed
by them. Empty balls means unstable points, and full balls stable ones.

{\bf Figure 2:} Final aspect of the phase diagram showing the equilibrium
points at infinity. Only one integral curve is chosen
to each region in the diagram to avoid it to become entangled.

\end{document}